\documentclass[11pt]{article}
\usepackage{epsf}
\usepackage{graphicx}        % standard LaTeX graphics tool

\def\ll{\label}

\def\c{\cite}

\def\r1{(\ref{$1})}

\def\ba{\begin{array}{c}}

\def\ea{\end{array}}

\def\l{\left}
\def\l({\left(}
\def\r){\right)}
\def\r{\right}

\def\la{\lambda}

 \def\be{\begin{equation}}
\def\bc{\begin{center}}
\def\ec{\end{center}}
\def\bit{\begin{itemize}}
\def\eit{\end{itemize}}
\def\ee{\end{equation}}
\def\ed{\end{document}}
\def\bea{\begin{eqnarray}}
\def\eea{\end{eqnarray}}
\def\efr{\end{flushright}}

\begin{document}
\title{{Exact accelerating solitons in   nonholonomic  deformation of the  KdV
equation with  two-fold  integrable hierarchy
}}
\author{ Anjan Kundu\\
 Theory Group \& CAMCS,
Saha Institute of Nuclear Physics\\
 Calcutta, INDIA\\
{anjan.kundu@saha.ac.in}} \maketitle
%\newpage
\noindent Short Title: {\it Solitons in  deformed  KdV 
and its two-fold integrable hierarchy  }
\\
\noindent {PACS:} 02.30.lk,
%integrable system
02.30.jr,
%PDE
 05.45.Yv,
%soliton
11.10.Lm,
%nonlin nonloc FT
\\
{\it Key words}: Nonholonomic deformation of  KdV,
 accelerating soliton, inverse scattering method,
  AKNS Lax pair, integrable hierarchies.
\vskip 1cm

\begin{abstract}
Recently proposed nonholonomic  deformation of the  KdV equation
% \c{6kdv}
 is  solved through inverse scattering method by constructing AKNS type Lax pair.
Exact N-soliton solutions
are found   for the basic field and the deforming function
showing  unusual ac(de)celerated motion. Two-fold integrable hierarchy  
is revealed, 
one with usual higher order dispersion and the other
with  novel
higher nonholonomic   deformations.
\end{abstract}

%45
\section{Introduction}
\label{sec1}

Nonholonomic constraints on field models are receiving increasing attention in
recent years \c{nholn}. As a  notable  achievement such 
 nonholonomic  deformation
 (NHD) has been  applied  to an integrable system, namely to the KdV equation 
preserving its integrability 
\c{6kdv}. For this  system, equivalent to a 6th order KdV equation,
   certain particular traveling wave
solutions are found, a linear problem is formulated and  several conjectures
 on important
issues  are put forward in \c{6kdv}.
The main expectations  for this   KdV equation with nonholonomic constraint  
  are: i) existence of an infinite set of higher conserved quantities,  ii) formulation of the Lax pair 
iii) application of the inverse scattering method (ISM), 
iv)
N-soliton solutions for the basic and the deforming field, v) elastic nature of 
soliton scattering  etc.  
Among these  conjectures  only the first one showing the
 existence of
an integrable hierarchy with   usual higher dispersion
 is proved recently \c{kup08}. Our aim here is to establish the rest with
explicit result.
%%%%%%%%%%

In particular we construct  AKNS type matrix Lax pair for
 this NHD of the
KdV equation, revealing an important connection between the 
 time evolution of the Jost function and the NHD
 of the nonlinear equation. Applying subsequently the ISM    
 we find the exact N-soliton solutions for both the field and the deforming
 function of
 the deformed    KdV, which  exhibit  the usual property  
of   soliton scattering, but  with an unusual accelerating
(or decelerating) soliton motion. Finally we unravel  a novel two-fold integrable
hierarchy for this system,
 one with the usual higher  dispersion \c{kup08} 
and the other
yielding a new type of deformed KdV with 
increasingly higher order nonholonomic   constraints.

\section{  ISM for the deformed KdV }

Recently proposed \c{6kdv}
     NHD of the KdV equation   \bea u_t-u_{xxx}-6uu_x
&=&w_x, \ll{kdvs} \\  w_{xxx}+4u w_x+2u_x w &=&0, \ll{kdvEE}\eea
%REV1
can be written by eliminating  deforming function $w$,  also as a 6th 
order KdV equation for $v_x=u $:
 \be  (\partial^3_{xxx}+4v_x \partial_x+2v_{xx})(v_t-v_{xxx}-3
v^2_x)=0.  \ll{6kdv}\ee
\subsection{AKNS type Lax pair}
We intend to find the  exact N-soliton solutions to the deformed KdV
equation  (\ref{kdvs}-\ref{kdvEE}) for the field $u $ and the deforming
function $w$ by the ISM, for which   
we construct first the AKNS type Lax pair  $U(\lambda), V(\lambda) $ to formulate the linear problem
 $\Phi_x=U\Phi, \ \Phi_t=V\Phi  $.
  We observe remarkably, that
such a Lax pair for the deformed KdV 
%(\ref{kdvs}-\ref{kdvEE}) 
 can be build up from the known pair 
 $U_{kdv}(\lambda),
V_{kdv}(\lambda) $ for the standard KdV equation \c{solit1}, by {\it deforming } 
only its time-Lax operator: 
\be U(\lambda)= U_{kdv}(\lambda), \ \
 V(\lambda)=V_{kdv}(\lambda)+V_{def}(\lambda), \ll{LAdef}\ee 
Where the deforming  operator is given as
\be 
V_{def}(\lambda)=\frac 1 2( \lambda^{-1}G^{(1)}+ \lambda^{-2}G^{(2)})
\ll{Vdef} \ee
with 
\bea 
G^{(1)}&=&i w \sigma^3- w_x \sigma^+\nonumber \\
G^{(2)}&=& \frac { w_x} 2 \sigma^3+ i w \sigma^-+ e\sigma^+
, \ e_x=iu w_x.
\ll{G12} \eea
%REV2
Here $u(x,t) $ is the KdV field and  $w (x,t) $ is the deforming function  with its
 asymptotic  $\lim_{|x | \to \infty}w=c(t) $,  being an arbitrary
function in time.
%REV3
 We can check that
 the flatness condition  $U_t-V_x+[U,V]=0 $ of the Lax pair
(\ref{LAdef})
 yields the deformed  KdV   equation (\ref{kdvs}-\ref {kdvEE}), where     
the undeformed
part is given by  the standard pair \c{solit1}
\bea U_{kdv}(\lambda)&=&i (\lambda\sigma^3
+U^{(0)}), \ \ U^{(0)} =u(x,t)\sigma^++ \sigma^- \ll{Ukdv} \\
V_{kdv}(\lambda)&=&
iU^{(0)}_{xx}-4i\lambda^3\sigma ^3+2 \sigma ^3 \lambda (
-U^{(0)}_x+i(U^{(0)})^2)   -4i  U^{(0)}\lambda^2  \nonumber \\ 
&+&2i
(U^{(0)})^{3} -[ U^{(0)}, U^{(0)}_x], \ll{Vkdv}\eea
while the deformed part with the nonholonomic   constraint 
is generated  by the new addition  (\ref{Vdef}).
Therefore we may  draw an intriguing conclusion  that   the NHD of the
KdV  (\ref{kdvs}-\ref {kdvEE}) can  be   linked to the deformation
in the time evolution of the Jost function, which is  related to $V_{def}(\lambda) $ as
given by     (\ref{Vdef}-\ref {G12}). We  see in the sequel that this
fact leads to an unusual solitonic property   in the deformed
KdV, namely the  possibility of   ac(de)celerated soliton   motion. 

\subsection{Exact soliton solutions}
For the deformed KdV equation (\ref{kdvs}-\ref {kdvEE}) 
 no exact solution, except  a few particular solutions, could  be found
\c{6kdv}. We derive  for the same equation  exact N-soliton
solutions, which is a clear signature of  
complete integrability   of a nonlinear system.      
It is important to note that,  the evolution of the basic KdV field
$\partial_tu $ is sustained  here from two different  sources:
$\partial_{t_0}u=\partial_{t}u |_{c(t) =0} $ and $
\partial_{t_d}u=c(t)\partial_{\tilde c(t)}u$. The first one is generated by
the standard dispersive and nonlinear terms in  equation (\ref{kdvs}), while the
second one is sourced by the deforming term $w_x $. Therefore the deforming
function  $ w$ satisfying the nonholonomic constraint (\ref{kdvEE}), can be
determined self-consistently  through the KdV field as
\be w(x,t)= c(t) \int u_{\tilde c(t)} dx +c(t), \ll{w-u} \ee
where the arbitrary function  $c(t)$ acting as a forcing term sitting at the space
boundaries  $ \lim_{x \to
\pm \infty}w(x,t)=c(t) $ arises as an integration constant. 

 Therefore we can find  by   applying 
  the ISM  the  exact soliton solutions  for
the basic as well as for the perturbing field,  interdependent on each other. 
Recall  that
 in using
  the ISM through the associated  linear problem, the space-Lax operator $U(\la ) $ describing the
scattering of the Jost functions, plays the key role.
 Only at the final stage we need to fix   the time
evolution of the solitons through  the time-dependence of the spectral data,
 determined  in turn  by the asymptotic value of the time-Lax operator   
$V(\la )$.
Note that since in the case of  the deformed  KdV equation the space-Lax operator
(\ref{LAdef}) is given by
the same operator as in the standard  KdV (\ref{Ukdv}), the steps for 
its ISM follows   the same initial
path as for the KdV equation \c{solit1}.
Therefore referring the readers to the original literatures for   details
 we produce  the explicit  form  of the N-soliton  for
the KdV field $u(x) $ as an exact solution to the deformed system
(\ref{kdvs}-\ref{kdvEE}),
 or equivalently  to  the 6th order KdV (\ref{6kdv})  as 
\be u_N(x)=2 \frac {d^2} {dx^2}\ln det A(x) \ll{Nsol}\ee
where  the matrix function $A(x) $ is expressed through its elements as
\be A_{nm}=\delta_{nm}+ \frac {\beta_n}
{\kappa_n+\kappa_m}e^{-(\kappa_n+\kappa_m)x} \ll{AN}\ee
Here  parameters $ \kappa_n, n=1,2,\ldots, N$, denote the
time-independent zeros   of the scattering matrix element
$a(\la=\la _n )=0 $, along the imaginary axis: $\la_n=i\kappa_n $  and   $
\beta_n(t)=b(\la = \la _n) $ are the time-dependent spectral data to be determined
from  $ V(\lambda)=V_{kdv}(\lambda)+V_{def}(\lambda) $,
at $ x \to \pm \infty$.
We notice that due to $u \to 0, w \to c(t) $ at $ x \to \pm \infty$,
  the asymptotic value of (\ref{Vkdv}):  $V_{kdv}(\lambda) \to 
-4i\lambda^3 \sigma^3
$ is the usual one, 
 while that of 
$V_{def}(\lambda) \to \frac i 2 \lambda^{-1}
c(t)\sigma^3  $ determines   the crucial effect of the deformation.
As a result we obtain 
\be \beta_n(t)=\beta_n(0) e^{-(8\kappa_n^3t -
 \frac {\tilde c(t)}{\kappa_n})}, \ \tilde c_t= c 
 \ll{bett}\ee 
determining finally the evolution of the soliton through (\ref{AN}).

The exact N-soliton solution for the deforming function $ w(x,t)$, induced through (\ref{w-u}) by
the solution   of the basic field (\ref{Nsol}), 
therefore can be given by 
\be w_N(x,t)=2 c(t) \frac {\partial^2} {\partial x \partial {\tilde c(t)}}(\ln det
A(x,t)) +c(t), \ll{Nsolw}\ee
where $ A(x,t)$ is the same matrix function  (\ref{AN}) with
its time-dependence (\ref{bett}).
 
To examine  the deforming effect on   solitons     in more
detail   we analyze  
 particular  cases of  solution (\ref{Nsol}) and (\ref{Nsolw}) for $N=1,2$.
 
1-{\it soliton} solution of  NHD of the KDV
 equation as reduced from  (\ref {Nsol}-\ref{Nsolw})
     can be expressed as
\bea u_1(x,t)&=&\frac {v_0} 2 {\rm sech}^2 \xi,  \ \  \  \xi= \kappa(x+vt)+\phi,  \ll{1s} \\ 
 w_1(x,t)&=&c(t)(1- {\rm sech}^2
\xi), \ll{1sw} \eea 
with $  v=v_0+v_d$, where $v_0=4\kappa^2 $ is the usual constant KdV soliton  velocity, while 
$v_d=- \frac
{2\tilde c(t)}{v_0 t}$ is its unusual time dependent part 
induced by the deformation. We stress again that the deforming function is
determined by the dynamics of the field $u$, which in turn is forced 
self-consistently  by
the deforming  field. Inserting the explicit  soliton solutions (\ref{1s},
\ref{1sw}) for both $u $ and $w$
in the nonholonomic deformation of the KdV (\ref{kdvs}-\ref{kdvEE}) 
one can  directly check  the validity of these exact solutions. Notice that
the
time-dependent asymptotic value of  the deformation: $c(t) $  acts here 
like a forcing term sitting  at the space boundaries, which  for $c(t)=c_0 t $  with
$c_0>0$ forces the soliton  to accelerate, while with $c_0<0 $ 
makes the soliton to decelerate and finally revert its direction (see Fig. 1).
 It can
also be noted that 
less  the original soliton velocity $v_0 $, more is  the deforming velocity $
v_d$, which is physically consistent since the forcing  term in
general  must have more
prominent  effect on 
slow moving solitons.

The exact  2-{\it soliton} in  the deformed  KdV
  can be  derived similarly from 
(\ref {Nsol}-\ref{Nsolw}) with $N=2 $ in the explicit form:  
\bea u_2(x,t)&=& \frac {2} {D^2}(D_{xx}D-D_x^2), \quad \  D=1+e^{-\xi_1}+e^{-\xi_2}+
p_{12}e^{-(\xi_1+\xi_2 )} \ll{2su} \\  
w_2(x,t)&=& c(t)\left ( 1+ \frac {2} {D^2}(D\tilde D_{x}-\tilde D D_x) \right ),
 \ \tilde D= \frac 1 {\kappa_1}e^{-\xi_1}+\frac 1 {\kappa_2} e^{-\xi_2}+
p_{12}(\frac 1 {\kappa_1} +\frac 1 {\kappa_2} )e^{-(\xi_1+\xi_2 )}
 \ll{s2}\eea
where the scattering amplitude $p_{12}=\left(\frac {\kappa_1-\kappa_2} 
{\kappa_1+\kappa_2} \right)^2$
and  $\xi_n=2 \kappa_n(x+v_nt)+\phi_n, \ n=1,2 $ with $ v_n=v_{0n}+v_{dn}$.
The usual constant soliton velocities $v_{0n}=4\kappa_n^2, n=1,2 $
 of the KdV equation is boosted
here by time-dependent  velocities $v_{dn}=- \frac
{2\tilde c(t)}{v_{0n}t}, n=1,2  $, caused by the nonholonomic deformation. The  
 scattering of solitons  for the field $ u$ as described by the solution
(\ref{2su}) with $\tilde
c(t)=\frac {c_0} 2 t^2 $ 
is depicted in Fig. 2, which shows  the
usual elastic collision of  solitons as conjectured in \c{6kdv}, but
  with an unusual dynamics due to the accelerating  motion of the solitons.
\begin{figure}[c]
\epsfxsize=0.96\textwidth

\qquad \qquad \qquad \qquad \ \ \ \ \ 
\includegraphics[width=6.cm,height=5.1 cm]{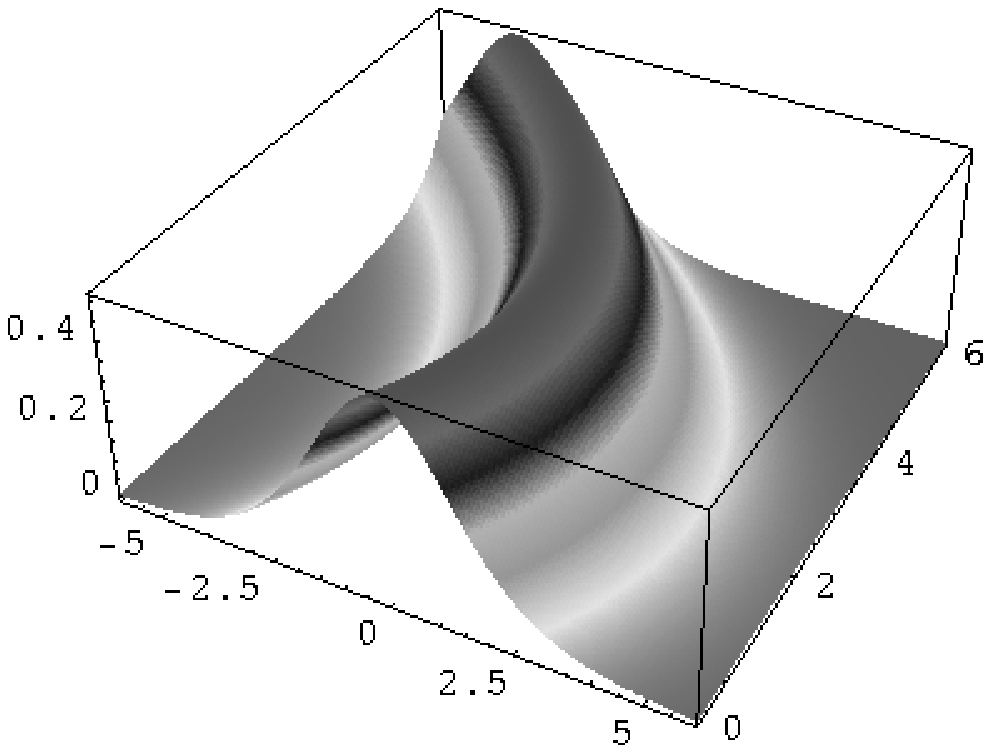}
%{s1skdvF.eps}
 
%\includegraphics[width=4.cm,height=3.1 cm]{sgd11s.eps}
 
\caption{   Exact soliton solution $ u_{1}(x,t)$ of  the KdV  field   
for the nonholonomically deformed equation (\ref{kdvs}-\ref{kdvEE}). 
%or equivalently of the 6th order KdV (\ref{6kdv}),
%integrable nonholonomic 
%deformation of the  KdV equation (\ref{})with
  showing usual localized form of the soliton, but  with its 
unusual decelerating motion, as evident from the bending of soliton
trajectory with time.
% dynamics for a) 
% The original
%soliton velocity $v_0 $  is boosted by  an additional
%$v_d= - \frac c {v_0} $ for constant $c $ due to   the source
%riving. If $ c=\alpha t $ the soliton can accelerate (decelerate)
%or +ve (-ve) $ \alpha$.
}
%\end{center}
\end{figure}
\begin{figure}[c]
\epsfxsize=0.96\textwidth

\qquad \qquad \qquad \qquad \ \ \ \ \
\includegraphics[width=6.cm,height=5.1 cm]{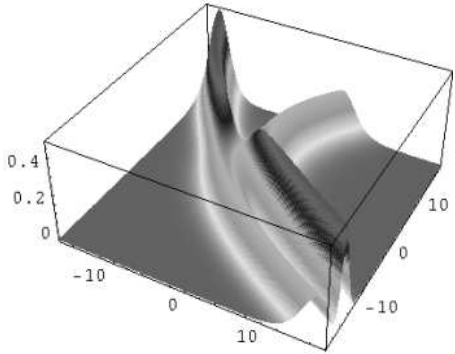}
%{s2skdvF.eps}
 
%\includegraphics[width=4.cm,height=3.1 cm]{sgd11s.eps}
 
\caption{Exact 2-soliton solution 
$ u_2(x,t)$ of  the KdV  field   
for the deformed equation (\ref{kdvs}-\ref{kdvEE}).
 or equivalently for (\ref{6kdv}). Usual elastic 
soliton scattering with phase shift is evident, though   the 
 dynamics  here is dominated by their unusual accelerating  motion, reflected 
in the bending of soliton trajectories.
}
%\end{center}
\end{figure}

\section{Two-fold integrable hierarchy for the deformed KdV equation}
The well known integrable hierarchy of the standard KdV equation is given
by \c{solit1} $u_t=B_1(\frac {\delta H^{kdv}_{n+1}} {\delta u})=
B_2(\frac {\delta H^{kdv}_{n}} {\delta u}) $,
with $B_1=\frac {\partial} {\partial x} $ and 
$B_2=\partial^3 +2(u\partial+\partial u) $ ,
 where $H^{kdv}_n,
n=1,2,\ldots $  are the higher Hamiltonians of the KdV hierarchy, e.g.
$H^{kdv}_1=u, \ H^{kdv}_2=\frac {u^2} 2, H^{kdv}_3=\frac 1 3 {u^3} -\frac 1 2 u^2_x  $
 etc., with $n=2 $ yielding the KdV equation.
As has been shown in \c{kup08} an integrable  hierarchy with the same Hamiltonians
exists also for the deformed  KdV equation  with NHD     of the  
equations as
\be 
u_t=B_1(\frac {\delta H^{kdv}_{n+1}} {\delta u} -w), \ B_2(w)=0,
\ll{hi1}\ee    
 For $n=2 $ one obviously recovers
the known deformed KdV equation (\ref{kdvs}-\ref{kdvEE}).
This usual type of hierarchy with higher dispersions can be generated from
the AKNS Lax pair, where the space-Lax operator remains same  as the original
one (\ref{Ukdv}), but the time-Lax operator is changed to
$V(\la )=V_{kdv}^{(n)}(\la )+V_{def}(\la ) $. Here the deforming part 
$V_{def}(\la ) $ is the same as (\ref{Vdef}-\ref{G12}),
 while $V_{kdv}^{(n)}(\la ) $
is the higher  generalization of  (\ref{Vkdv}), where  polynomial 
in  spectral parameter $\lambda $  up to  n-th power appears. Such higher order
time-Lax operator can be constructed by expanding this matrix  in the powers
of $\lambda $ and determining the matrix coefficients recursively from
the flatness condition, as done in the standard AKNS treatment \c{solit1}. Another simpler solution for
this problem 
based on the dimensional analysis and  identification of 
 the building blocks of the Lax
operators has  been
proposed recently \c{kun08}.

We discover,   apart from the usual integrable hierarchy  given above,
an unusual  hierarchy for the nonholonomic deformation  of the KdV equation, which can be 
represented   by the same 
deformed KdV (\ref{kdvs}) but with  higher order nonholonomic constraints on
the deforming function $w $. This novel integrable hierarchy can be
generated as the flatness condition of a Lax pair, 
where the space-Lax operator  remains   as  (\ref{Ukdv}), while
 in the time-Lax operator only  the deforming part  changes as 
$V(\la )=V_{kdv}(\la )+V_{def}^{(n)}(\la ) $. Unlike the above KdV
hierarchy,
 $V_{def}^{(n)}(\la )=\frac 1 2\sum_j^n \lambda ^{-j}G^{(j)}, j =1,2, \ldots n$
 contains only  
 $-ve $ powers of $\lambda$ up to $n $,  with $n $-number of   deforming  matrix
coefficients $G^{(j)} $. 
The consistency condition generates this 
new integrable hierarchy of nonholonomic   deformations 
 given by the same deformed evolution  equation (\ref{kdvs}), but where    constraint 
(\ref{kdvEE}) is generalized now to n-th order through 
 a set of coupled differential
equations:
  \bea
 u_t-u_{xxx}-6uu_x&=& G^{(1)}_{12}, \nonumber \\
 G^{(1)}_x&=&i[U^{(0)},G^{(1)}]+i[\sigma_3, G^{(2)}], \nonumber \\ \ \ldots \ & & \ \ldots \
 ,
\nonumber \\G^{(n-1)}_x&=&i[U^{(0)},G^{(n-1)}]+i[\sigma_3, G^{(n)}],
\nonumber \\ G^{(n)}_x&=&i[U^{(0)},G^{(n)}].
 \ll{EEn} \eea
 Clearly this hierarchy  reduces to   NHD of the KdV
(\ref{kdvs}-\ref{kdvEE}) for $
 n=2$, with deforming operator  $V_{def}^{(2)}(\la )$ reducing 
to (\ref{Vdef}-\ref{G12}).

%SHESH3004

\section{Concluding remarks}

We prove here a number of conjectures on the 
recently proposed 
  nonholonomic deformation    of the KdV equation, unraveling  its
 several  unexpected features.   
In particular we construct  AKNS type matrix Lax pair for
 this deformed
KdV equation, showing  an intriguing  connection between the 
deformation of the  time evolution in the associated linear problem 
 and the deformation
 of the nonlinear equation. Applying  the inverse scattering method     
 we find exact N-soliton solutions for the basic as well as the deforming
 field for the deformed KdV equation, which gives  also the solution of  
 the 6th order    KdV equation. Such solitons  exhibit in spite of  the
isospectral flow   an unusual accelerated motion,
 which  is however consistent  with the particle motion under force.
The  deforming function  $w(x,t) $, as seen from (\ref{hi1}), enters  in the
original hierarchy 
 of the  KdV equation  as a perturbation together
with   a nonholonomic differential constraint on it.
 The driving term 
sitting at the space-boundaries: $c(t)=w(\pm \infty , t) $,
which can  be  an arbitrary function in time, forces the field soliton to accelerate
or decelerate, with the perturbing soliton itself created by the field
soliton in a self-consistent way.

  We discover also an unique  two-fold integrable
hierarchy for this deformed system,
 one with  usual higher  dispersion found already    
and the other with new 
increasingly higher order nonholonomic  deformation. 
Extension of nonholonomic deformation to other integrable models like NLS,
sine-Gordon, mKdV etc. is under investigation \c{kun08}.
 
%SHESH1004

 \end{document}